\documentstyle[12pt,epsf]{article}
\begin{document}
\title{\bf Series expansion and computer simulation studies
of random sequential adsorption}

\author{Jian-Sheng Wang\\
Department of Computational Science,\\
National University of Singapore,\\
Singapore 119260, Republic of Singapore}

\date{8 March 1999}

\maketitle
\begin{abstract}
We discuss two important techniques, series expansion and Monte
Carlo simulation, for random sequential adsorption study.
Random sequential adsorption is an idealization for surface
deposition where the time scale of particle relaxation is much
longer than the time scale of deposition.  Particles are
represented as extended objects which are adsorbed to a
continuum surface or lattice sites.  Once landed on the surface,
the particles stick to the surface.  We review in some details
various methods of computing the coverage $\theta(t)$ and
present some of the recent and new results in random sequential
adsorption.
\end{abstract}

\section{Random sequential adsorption problem}
    
\begin{figure}[tb]
\center{\leavevmode\epsfxsize=0.7\hsize\epsfbox{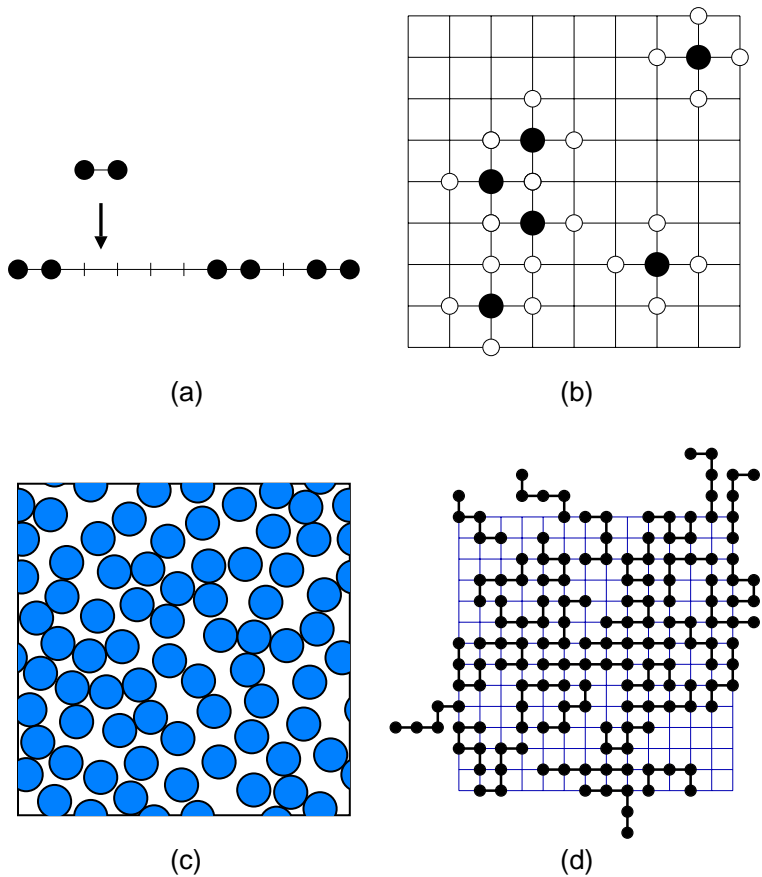}}
\caption[RSA config]{Some typical random sequential adsorption
configurations. (a) Dimer deposition on a one-dimensional lattice;
(b) monomer with nearest neighbor exclusion; 
(c) the jamming state of hard discs on continuum; 
(d) self-avoiding random walks of length $N=5$ on square lattice.}
\label{fig:config}
\end{figure}

The random sequential adsorption (RSA) problem arises from a
number of situations in physics, chemistry, and biology.  One of
the early RSA problem comes from polymer chemistry.  Flory
\cite{flory} was interested in the dimerization of polymer chain.
The adjacent monomers on the side of the chain form dimers one
at a time until no pair of monomers is available.  He found the
final reacted monomers based purely on geometric argument.  It
turns out that this problem is equivalent to a random sequential
adsorption of dimers on a one-dimensional lattice.  Consider the
one-dimensional lattice, with all sites empty initially.  At time 
$t > 0$, dimers, each of which occupies two consecutive sites, are
dropped on the lattice at a rate $k$ per second equally likely
at any locations (see Fig.~\ref{fig:config}(a)).  As long as the
two sites are empty, a dimer lands on the lattice.  If any one
of the two sites is already occupied, the deposition is
rejected.  This is the basic RSA model.

RSA of many different geometric objects has been studied.  The
deposition of unit line segments on a one-dimensional continuum
is called car-parking problem \cite{renyi}.  The deposition of
discs on two-dimensional continuum \cite{feder} is relevant to
the adsorption of colloids or protein molecules on glass surface
\cite{finegold-donnell,onoda-liniger}.  Lattice models of various
geometries are studied.  In Fig.~\ref{fig:config} we show some
typical configurations of RSA models.

The very basic question of RSA is the time dependence of the
coverage, $\theta(t)$, and the jamming coverage,
$\theta(\infty)$.  The approach to the jamming coverage is also
of considerable interest.  It turns out that the approach to the
jamming coverage is to some extent universal.  For lattice
models, the approach to jamming is always exponential due to the
discreteness of the problems \cite{privman-wang-nielaba-PRB},
while for continuum problems the asymptotic dependences exhibit
power laws which depend on the symmetry of the deposited objects
\cite{pomeau,swendsen,viot-tarjus-europhys-lett,tarjus-viot-PRL}.

A comprehensive review on RSA and cooperative sequential
adsorptions is given by Evans \cite{evans-review}.  Bartelt and
Privman \cite{bartelt-privman-review} reviewed, among other
things, mean-field approximations.  Progress on experiment work
is given by Ramsden \cite{ramsden}.  Random sequential adsorption has
many aspects and is rich with problems.  In this article, we
limit our scope to consider two important approximate techniques
for computing the coverage of RSA.  They are the series
expansion and Monte Carlo simulation.  The emphasis is on
methods rather than specific results. Only one-dimensional
problems
\cite{page,mackenzie,cohen-reiss,gonzalez-hemmer,bonnier-boyer} 
and quasi-one-dimensional problems 
\cite{evans-tree,baram-kutasov-1d,fan-percus-ladder} have exact solutions.
Two-dimensional problems are most likely intractable.  Because
of this, the approximate methods like series expansion and Monte
Carlo are very useful for the study of RSA.  We present some of
the technical details which are usually given only briefly in
original research papers.  With examples, we show how the series
can be obtained and analyzed with Pad\'e approximation.  For
Monte Carlo simulation, the emphasis is on efficient algorithms
for RSA simulation.

\section{Series expansions}

One of the well-developed methods in studying random sequential
adsorption is series expansion around time $t=0$.  The method
gains experience from series expansion studies of other problems
in condensed matter physics \cite{guttmann}.  Series expansion
approach was the only successful method to obtain nonclassical
results in the study of critical phenomena for a while before
the advent of renormalization group method \cite{domb}.  The
methods and technique developed, especially the Pad\'e
approximation technique \cite{baker}, have been also applied
successfully to random sequential adsorption.  The technique
offers a systematic approach with controlled approximation.
Unlike the approach with truncated rate equations
\cite{evans-nord-JSP85,nord-evans-JCP85,schaaf-talbot-rabeony-reiss}, 
series expansion can be automated on computer relatively easily,
and the number of terms obtained can be rather high.  In many
cases, the accuracy is higher than or comparable to most accurate
Monte Carlo simulation results.

The series expansion for the rate of adsorption as a function
of particle density has been introduced by Widom
\cite{widom}.  It was shown that this type of expansion is
similar to virial expansion for the equation of state.  In fact,
the first two coefficients involving one and two-particle
distributions are the same.  The difference between equilibrium
system and RSA starts at third order.  Series for hard discs on
two-dimensional continuum up to third order were first obtained
by Schaaf and Talbot \cite{schaaf-talbot}.  Series expansions
for lattice models and continuum systems have been considered by
Hoffman \cite{hoffman} and Evans \cite{evans-series}.  Evans
presented rate equation approach for deriving series.  The
series in time and in density are related by a transformation, so
they are equivalent.  High-order series with the help of
computer were derived by Baram and Kutasov \cite{baram-kutasov},
Dickman {\sl et al} \cite{dickman-wang-jensen}, Bonnier {\sl et
al}
\cite{bonnier-series}, Baram and Fixman \cite{baram-fixman}, and
Gan and Wang
\cite{gan-wang-JCP}.

\subsection{Rate equations}

The RSA process can be described fully with a master equation
of the form 
\begin{equation}
  { \partial P(\{\sigma\}, t) \over \partial t}  = 
 \sum_{ \{\sigma'\}} 
 \Gamma(\{\sigma\}, \{\sigma'\}) P(\{ \sigma'\}, t),
\label{eq:master}
\end{equation}
where $\{ \sigma\}$ is a set of state variables which completely
specify the state of the system; $P(\cdots)$ is the probability
of such a state; and $\Gamma$ is a transition matrix.  For
notational convenience, we shall consider discrete lattice
models.  Continuous space problem can be generalized easily.
Then $\{\sigma\}$ denotes a set of occupation variables
$\sigma_i$ located at site $i$.

Although Eq.~(\ref{eq:master}) contains the complete information
of the process, it is rather difficult to deal with efficiently
due to the high dimensionality of the problem.  Thus, in most of
the analytic treatments, reduced quantities are worked with.
One of the most important such quantities is the marginal
probability distribution:
\begin{equation}
   P(G) = \!\!\!
   \sum_{\{\sigma\},\,\sigma_i = \sigma_i^0 {\ \rm for\ } i \in G} 
   \!\!\! P(\{\sigma\}),
\end{equation}
where the summation is over all the possible states such that a
given set $G$ of sites takes a known set of values.  That is,
$P(G)$ is the probability that a given set of sites having
specified values.  For the standard RSA problem, it is sufficient
to consider only a set of connected sites which are unoccupied.

\begin{figure}[bt]
\center{\leavevmode\epsfxsize=0.7\hsize\epsfbox{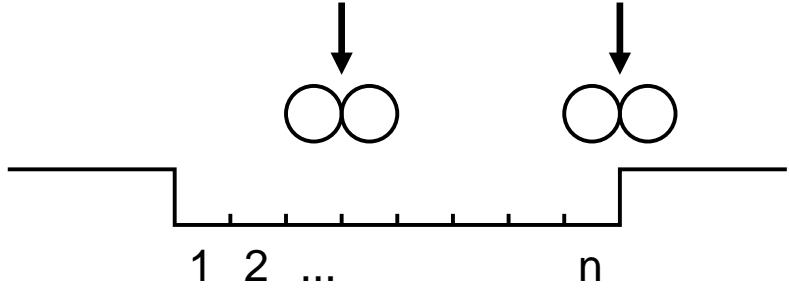}}
\caption[Rate eqn]{The one-dimensional dimer deposition.
The sites 1 to $n$ are known to be empty, but the status of other
sites are unknown.}
\label{fig:rate}
\end{figure}

Our task is to write down a set of differential equations for
$P(G)$.  In principle, we can derive them from the master
equation (\ref{eq:master}).  This is not really necessary, and
we can give the equations based on physical meaning of $P(G)$.
Let us take a look of the equations associated with a random
sequential deposition of dimers on a one-dimensional lattice.
In this case, we consider the probability $P_n$ that a
consecutive $n$ sites are empty.  Assuming translational
invariance of the initial conditions, this probability does not
depend on the specific locations.  Due to the simple geometry in
one dimension, the set $G$ of $n$ connected sites being empty
can be characterized by a single integer $n$.  Let us consider
the change of $P_n$ by the deposition process.  We assume that
the $n$ empty sites are at $i=1,2,\ldots,n$.  Clearly,
depositions of dimers involving the sites $i \leq 0$ or $i > n$
do not change $P_n$, while depositions inside empty sites on 1
to $n$ destroy the consecutive empty sequence, the probability
of which is proportional to $P_n$ for each of the $n-1$ possible
ways of depositions.  At the two ends, the empty sequence can be
destroyed with one site at 0 (or $n+1$) outside the sequence and
the other site at 1 (or $n$) in the sequence, see
Fig.~\ref{fig:rate}.  In order for this to happen, the site just
outside the considered sequence should be empty, the probability
of which is $P_{n+1}$.  There are two ways to do this.  Putting
all these together, we have,
\begin{equation}
   - k^{-1} {d P_n(t) \over dt} =  
         (n-1) P_n(t) + 2 P_{n+1}(t), \qquad n=1, 2, 3, \cdots.
\end{equation}
The proportionality constant $k$ sets the time scale.  Without
loss of generality, we can take it to be unity.  This equation
can be solved \cite{cohen-reiss} to give
\begin{equation}
 P_n(t) =  \exp[ - (n-1) t - 2 + 2 e^{-t}].
\label{eq:1dexact}
\end{equation}
Note that the coverage is just $\theta(t) = 1-P_1(t)$.

Similar consideration gives so-called rate equations
\cite{vette} for other deposition processes.  For example, 
in deposition of dimers on square lattice, the basic quantity of
interest is the probability $P(G)$ that the given set $G$ of
sites are empty; we do not care other sites being occupied or
empty.  The rate of decrease is proportional to the probability
that the current configuration $G$ can be destroyed by
depositing a dimer with at least one site in $G$.  Thus we have
\begin{equation}
- { d P(G) \over dt } =\mskip-24mu
\sum_{{\rm ways\ of\ destroying\ } G}\mskip-30mu  P(G'), 
\end{equation}
where the summation runs over all possible ways of depositing a
dimer at a pair of empty sites such that at least one site is in
$G$; $G'=G$ if two of the dimer sites of a deposition attempt
are within $G$, or $G'$ is one site more than $G$ so that a
deposition with one site in $G$ and one site outside $G$ can be
carried out.  The first two equations look like these:
\begin{equation}
- { dP({\rm o})\over dt }  = 4\,P({\rm oo}), 
\end{equation}
\begin{equation}
- { dP({\rm oo})\over dt } = P({\rm oo}) + 2\,P({\rm ooo}) 
+ 4\,P(\vbox{\hbox{\lower 7pt \hbox{o}}\hbox{oo}}). 
\end{equation}
There are four ways to destroy a single empty site, provided
that the nearest neighbor site is also empty.  Using the
assumption that initial conditions are lattice symmetry
invariant, we can write them simply as $4 P({\rm oo})$.  The
second equation is derived similarly.

For a general discussion, we write the rate equations
symbolically as
\begin{equation}
 { dP(G)\over dt}  = {\cal L} P(G),
\label{eq:rate}
\end{equation}
where $\cal L$ is a linear operator defined by
\begin{equation}
{\cal L} P(G) = \sum_{G'} c_{G'} P(G').
\end{equation}
The $n$-th derivative is then 
\begin{equation}
{ d^n P(G)\over dt^n } 
= {\cal L}^n P(G), \qquad {\rm with\ }\quad P(G)\big|_{t=0} = 1 
\quad {\rm\ for\ all\ }G.
\end{equation}
We assume that all sites are empty at $t=0$.

\subsection{Simple counting methods for lattice models}

We can obtain the rules for series expansion starting from the
rate equation, Eq.~(\ref{eq:rate}).  We begin with a concrete
example of the random sequential adsorption of single site
occupation with nearest neighbor exclusion on a square lattice.
A site can be occupied by a particle, as long as the four
nearest neighbors are empty. Figure~\ref{fig:config}(b) shows a
sample configuration of the nearest neighbor exclusion RSA
model.

The first few rate equations read:
\begin{equation}
 - { d P(\hbox{o}) \over dt } =  P(\vbox{\hbox{\hphantom{o}o}\vskip 
-7pt\hbox{ooo}\vskip -7pt\hbox{\hphantom{o}o}}), 
\label{eq:rate1}
\end{equation}
\begin{equation}
 - { d \> \over dt }  P(\vbox{\hbox{\hphantom{o}o}\vskip 
-7pt\hbox{ooo}\vskip -7pt\hbox{\hphantom{o}o}})= 
   P(\vbox{\hbox{\hphantom{o}o}\vskip 
-7pt\hbox{ooo}\vskip -7pt\hbox{\hphantom{o}o}}) 
 + 4 P(\vbox{\hbox{\hphantom{o}oo}\vskip 
-7pt\hbox{oooo}\vskip -7pt\hbox{\hphantom{o}oo}}), 
\label{eq:rate2}
\end{equation} 
\begin{equation}
 - { d \> \over dt } P(\vbox{\hbox{\hphantom{o}oo}\vskip 
-7pt\hbox{oooo}\vskip -7pt\hbox{\hphantom{o}oo}}) = 
 2 P(\vbox{\hbox{\hphantom{o}oo}\vskip 
-7pt\hbox{oooo}\vskip -7pt\hbox{\hphantom{o}oo}})  
 + 2 P(\vbox{\hbox{\hphantom{o}ooo}\vskip 
-7pt\hbox{ooooo}\vskip -7pt\hbox{\hphantom{o}ooo}})  
 + 4 P(\vbox{\hbox{\hphantom{o}o}\vskip -7pt\hbox{ooo}\vskip 
-7pt\hbox{oooo}\vskip -7pt\hbox{\hphantom{o}oo}}),  
\label{eq:rate3}
\end{equation} 
where a pattern with o's denotes a set of sites of that
geometric arrangement which are empty (while other sites are
unspecificed).  We have taken into account the translational and
rotational symmetry of the problem.  The equations are
genenerated as follows.  For each given configuration $G$, we
consider all possible ways of destroying the configuration by a
deposition of a particle at each one of the empty sites. If the four
neighbors are already known to be empty, the rate is simply
proportional to the original probability $P(G)$.  If the site is
adjacent to sites of unknown status, we require those sites to
be empty.  Thus this second type of process creates new
configurations $G'$ which have more empty sites than $G$.

From this set of rate equations we can get the $n$-th derivative
of the probability that the sites in the set $G$ are empty,
$P^{(n)}(G) = d^n P(G)/dt^n|_{t=0}$.  With an initial condition
of empty lattice at $t=0$, the zeroth derivative $P(G,t=0)$
equals 1 for all configurations $G$. The most relevant one is
the probability that a single site is empty, $P(\hbox{o},t)$,
since $\theta(t) = 1 - P(\hbox{o},t)$ is the coverage.  Power
series expansion is obtained from the derivatives evaluated at
$t=0$:
\begin{equation}
  P(\hbox{o}, t) = \sum_{n = 0}^{\infty} 
                             P^{(n)}(\hbox{o}) { t^n \over n!}. 
\end{equation}
For the nearest neighbor exclusion model, we have 
$P(\hbox{o}) = 1$.  $P'(\hbox{o}) = -1$ from Eq.~(\ref{eq:rate1}).
From Eq.~(\ref{eq:rate1}) and (\ref{eq:rate2}), we get
$P''(\hbox{o}) = 
 -  P'(\vbox{\hbox{\hphantom{o}o}\vskip 
-7pt\hbox{ooo}\vskip -7pt\hbox{\hphantom{o}o}})  = 
   P(\vbox{\hbox{\hphantom{o}o}\vskip 
-7pt\hbox{ooo}\vskip -7pt\hbox{\hphantom{o}o}}) 
 + 4 P(\vbox{\hbox{\hphantom{o}oo}\vskip 
-7pt\hbox{oooo}\vskip -7pt\hbox{\hphantom{o}oo}}) = 5$. 
Similarly, we find
$P'''(\hbox{o}) = 
 -  P''(\vbox{\hbox{\hphantom{o}o}\vskip 
-7pt\hbox{ooo}\vskip -7pt\hbox{\hphantom{o}o}})  = 
   P'(\vbox{\hbox{\hphantom{o}o}\vskip 
-7pt\hbox{ooo}\vskip -7pt\hbox{\hphantom{o}o}}) 
 + 4 P'(\vbox{\hbox{\hphantom{o}oo}\vskip 
-7pt\hbox{oooo}\vskip -7pt\hbox{\hphantom{o}oo}}) = -37$.  
Thus, to third order, the expansion is
\begin{equation}
P(\hbox{o},t) = 1 - t + { 5 \over 2} t^2 - { 37 \over 6} t^3 + O(t^4).
\end{equation}

The process of obtaining the answer can be thought as counting
the number of patterns in $n$ generations.  We note that for
each current configuration $G$, we go over the sites of $G$
once, each generated a next generation pattern, which may or may
not be the same as the parent pattern.  The expansion
coefficient $S(n) = (-1)^{(n+1)} d^{n+1} P(\hbox{o},t)/dt^{n+1}
|_{t=0}$ is simply the total number of possible patterns (or
sequences), where two-dimensional coordinate $x_0 =(0,0)$ is
fixed at the origin, while $x_1$ varies in a domain $D(x_0)$,
and $x_2$ in a union domain $D(x_0)
\cup D(x_1)$, \dots, and $x_n$ varies in a domain $D(x_0) \cup
D(x_1) \cdots \cup D(x_{n-1})$.  The domain $D(x)$ is the center
site $x$ plus four nearest neighbor sites of $x$.  We write
\cite{dickman-wang-jensen}
\begin{equation}
   S(n) = \sum_{x_1 \in D(x_0)}\; \sum_{x_2 \in D(x_0) \cup D(x_1)} 
\cdots \sum_{x_n \in D(x_0) \cup D(x_1) \cdots \cup D(x_{n-1})} 1.
\label{eq:Sn}
\end{equation}
$S(0)$ is defined by $-dP(\hbox{o},t)/dt|_{t=0}$. 
By definition, $S(n)$ is always positive. 
In terms of $S(n)$, the expansion for the rate of adsorption is
\begin{equation}
\phi = - { dP(\hbox{o},t) \over dt} = 
                    \sum_{n=0}^\infty S(n) { (-t)^n \over n!}.
\end{equation}
This notation is convenient, for Eq.~(\ref{eq:Sn}) generalizes
to other type of lattice models by choosing a different domain
$D(x)$ and generalizes to random sequential adsorption on
continuum.  For anisotropic objects, $x$ should be understood as
position and orientation as well.

The easiest way of implementing this counting method is to use
recursive function calls.  Consider the following C-like
pseudo-code:

\bigskip
\vbox{
{\parskip=0pt\obeylines
\def\t{\hskip 12pt}
\t{\bf RSA}($G$, $n$)            
\t$\{$
\t\t $S(n)\, +\!\!= |G|$;
\t\t {\bf if} $(n \geq N_{max})$ {\bf return};
\t\t {\bf for each} $(x \in G)$  $\{$ 
\t\t\t {\bf RSA}$\Big(G \cup D(x)$, $n+1\Big)$;
\t\t$\}$
\t$\}$
}
}
\bigskip
In this program, $G$ is a set of empty sites, which can be
represented by a list of coordinates of the sites; $G \cup D(x)$
is the union of the set $G$ and the set consisting of $x$ and
its four nearest neighbors; $|G|$ is the cardinality of the set.
The variable $G$ is local to each function call, while $S(n)$ is
global.  We assume that $S(n)$ is initialized to zero.  The
recursion stops after $N_{max}$ in depth, which gives us results
for $S(0)$ to $S(N_{max})$.  If we evoke the program by {\bf
RSA}$(G, 0)$, the derivatives $d^nP(G) /dt^n |_{t=0} = (-1)^n
S(n-1)$, $1 \leq n \leq N_{max}+1$, are computed.

Clearly, the above algorithm can be improved.  First of all, we
need not count a configuration if it is the same as the parent
configuration. We shall discuss this in detail later in Section 2.5.
The second improvement is to rewrite the recursive calls by
nonrecursive procedure.  Since the set $G$ always contains the
parent set where $G$ is derived, it is possible to use only one
list representing all the $G$'s (one for each level in the 
recursion), each ends at some point in the
list.  Of course, the bookkeeping will be more complicated.
None of the above suggestions will improve the speed of
calculation substantially.  In next section, we discuss perhaps
the most efficient algorithm \cite{gan-wang-PhysA} for the
series expansions on lattices.

\subsection{More advanced methods}

We note that in the simple counting algorithm, we did not use
the symmetry of the problem.  Two patterns related by
translation or rotation symmetry should have the same
derivatives, but this is not recognized.  In order to exploit
fully the symmetry of the problem, we use the technique of
dynamic programming, and use sophisticated data structures.  The
starting point is again the rate equation, Eq.~(\ref{eq:rate}).
It turns out that the best algorithm is to combine this method
with a simple counting algorithm.

\begin{figure}[tb]
\center{\leavevmode\epsfxsize=0.5\hsize\epsfbox{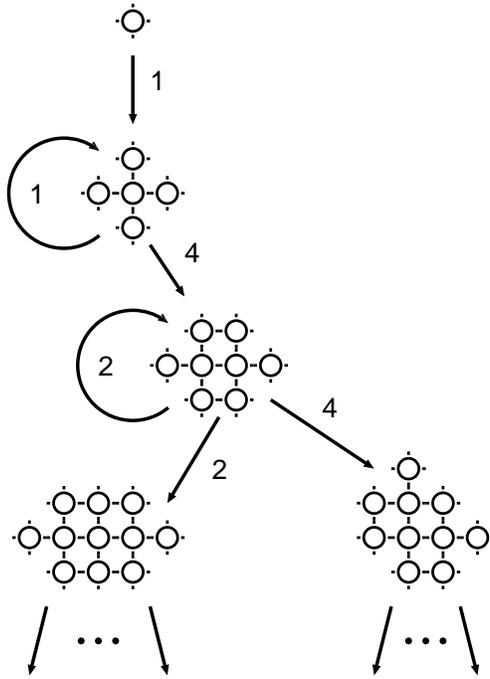}}
\caption[Expansion Tree]{The ``tree'' structure representing 
Eq.~(\ref{eq:rate1}) to (\ref{eq:rate3}) in the RSA series
expansion of the nomomer with nearest neighbor exclusion.  The
arrows point from parent to child generations; the numbers
indicate the expansion coefficients (number of equivalent
patterns).}
\label{fig:tree}
\end{figure}

A general function is coded which returns the right-hand side of
Eq.~(\ref{eq:rate}) when the configuration $G$, or the set $G$,
is given.  The set $G$ is represented as a list of coordinates
constructed in an ordered manner.  Note that each term on the
right-hand side of the equation is just the probability of the
set $G \cup D(x)$ being empty, where $x$ runs over the sites of
$G$.  Terms which are identical after translational and
rotational symmetry consideration are collected as one term with
associated coefficients.  Each rate equation is represented by a
node together with a list of pointers to other nodes.  Each node
represents a function characterized by the set $G$.  The node
contains pointers to the derivatives of this node obtained so
far, and pointers to the ``children'' of this node and their
associated coefficients, which form a symbolic representation of
the rate equations.  The computation of $n$-th derivative uses
the rate equations recursively.  Since each node is linked to
other nodes, the computation of the $n$-th derivative can be
considered as expanding a ``tree'' (with arbitrary number of
branches) of depth $n$.  The tree structure for the nearest
neighbor exclusion model is shown in Fig.~\ref{fig:tree}.
Unlike the standard tree, we allow loops linking back to earlier
generations.

The traversal or expansion of the tree \cite{cormen} can be done
in a depth-first fashion or a breadth-first fashion.  Each has a
different computational complexity.  A simple depth-first
traversal requires only a small amount of memory of order $n$.
However, the time complexity is at least exponential, $b^n$,
with a large base $b$.  A breadth-first algorithm consumes
memory exponentially, even after the number of the rate
equations has been reduced by taking the symmetry of the problem
into account.  The idea of dynamic programming can be
incorporated in the breadth-first expansion where the
intermediate results are stored and referred.  To achieve the
best performance, a hybrid of strategies is used to reduce the
computational complexity:

\begin{itemize}

\item Each configuration (pattern) is transformed into its
canonical representation; all configurations related by lattice
symmetry are considered as the same configuration.

\item We use breadth-first expansion to avoid repeated
computations involving the same configuration. If a
configuration has already appeared in earlier expansion, a
pointer reference is made to the old configuration.  Each
configuration is stored in memory only once.  However, storing
of all the distinct configurations leads to a very fast growth
in memory consumption.  For a quick check if a configuration is
already stored, we use the standard technique of hashing
\cite{cormen}.

\item The last few generations in the tree expansion use a
simple depth-first traversal to curb the problem of memory
explosion.

\item Parallel computation proves to be useful.
\end{itemize}

The program is controlled by two parameters $D$ and $C$.  $D$ is
the depth of breadth-first expansion of the tree.  When depth
$D$ is reached, we no longer want to continue the normal
expansion in order to conserve memory.  Instead, we consider
each leaf node afresh as the root of a new tree.  The
derivatives up to $(n-D)$th order are computed for this leaf
node.  The expansions of the leaf nodes are done in serial, so
that the memory resource can be reused.  The parameter $C$
controls the number of last $C$ generations which should be
computed with a simple depth-first expansion algorithm.  It is a
simple recursive counting algorithm, which uses very little
memory, and can run fast if the depth $C$ is not very large.  In
this algorithm the lattice symmetry is not treated.

This technique is used to obtain the RSA series for dimer and
nearest neighbor exclusion models on square and honeycomb
lattices \cite{gan-wang-JCP}, RSA with diffusional relaxation
\cite{gan-wang-PRE}, and Ising relaxation dynamics
\cite{wang-gan-PRE}.

\subsection{Series for continuum systems}

The results derived for lattice model can be generalized for
continuum system.  Consider the deposition of disc of diameter
$\sigma=1$.  One way to obtain the rules for series expansion in
continuum is to consider the limit of discretized lattice model.
The RSA of discs can be thought as approximate lattice problem
with a single-site deposition and exclusion of sites within some
distance of the occupied site.  The analog of Eq.~(\ref{eq:Sn})
is
\cite{dickman-wang-jensen},
\begin{equation}
   S(n)  =  \int\limits_{x_1 \in D(x_0)}\!\!\!\! dx_1 
\int\limits_{x_2 \in D(x_0) \cup D(x_1)}\!\!\!\!\!\!\!\! dx_2 \quad \cdots 
\int\limits_{x_n \in D(x_0) \cup D(x_1) \cdots \cup D(x_{n-1})} 
\!\!\!\!\!\!\!\!\!\!\!\!\!dx_n,
\label{eq:Sncont}
\end{equation}
where $x_i$ is interpreted as a $d$-dimensional vector, $dx_i$ is
a $d$-dimensional volume element, and $x \in D(x_i)$ is the set
such that $|x-x_i| \leq 1$.

The integral in Eq.~(\ref{eq:Sncont}) has a rather complicated
integration domain.  To simplify the integral, we introduce the
``Mayer function'' of a hard disc system,
\begin{equation}
f_{ij} = f(x_i - x_j) =  \cases{ -1, &  if $|x_i - x_j| \leq 1$;\cr
                         0, & otherwise.\cr}
\end{equation}
Then we can rewrite the integral with integration domain in the whole
space and with the integrand consisting of terms of products of
$f_{ij}$:
\begin{equation}
- S(1) = \int f_{01}\, dx_1,
\end{equation}
\begin{equation}
S(2) = \int f_{01}\, dx_1\int \Bigl[ f_{02} + f_{12}(1+f_{02}) \Bigr]\, dx_2,
\end{equation}
and in general 
\begin{equation}
(-1)^n S(n)  =  \int dx_1 \int dx_2 \cdots \int dx_n 
       \prod_{k=1}^n \left( \sum_{j=0}^{k-1} f_{jk} 
       \prod_{i=0}^{j-1} ( 1+f_{ik}) \right).
\label{eq:Snf}
\end{equation}
The result of expanding the terms leads to the following
graph-theoretic description: $(-1)^n S(n)$ is the sum of the
contributions from all connected $(n+1)$-point unlabeled
graphs; each term is the value of an integral associated with
the graph times an integer multiplicity factor.  The value of a
(labeled) graph is the integral of a form $\int\int \cdots \int
dx_1 dx_2 \cdots dx_n \prod f_{ij}$, called cluster integral.
The numerical factor is the number of topologically distinct
ways of labeling the graph, subject to the following
constraint: the labels are put down in sequential order from 0
to $n$, such that the current vertex $k$ is connected to a
vertex $j < k$.  There is a one-to-one correspondence between a
graph and an integral.  Each vertex $i$ of the graph corresponds
to an integration variable $x_i$ (except vertex 0), and each
edge $(i,j)$ of the graph corresponds to a function $f_{ij}$.
The definitions differ slightly from the normal convention
\cite{hansen}.  The diagrammatic rules give, for the particle
density, the following expression for the first few terms,
\begin{eqnarray}
\rho(t) & = & t + (\hbox{o--o}) {t^2 \over 2!} + \left( 2\; 
\vcenter{\vbox{\hbox{o}\vskip-7pt\hbox{\hskip1pt$|$}\vskip-7pt\hbox{o--o}}} + 
\vcenter{\vbox{\hbox{\hskip6.5pt o}\vskip-7pt\hbox{\hskip5pt$\bigwedge$}\vskip-7pt\hbox{o--o}}}
 \right) { t^3 \over 3!}  + \nonumber\\ &  &  \left( 2\;
\vcenter{\vbox{\hbox{o\phantom{--}o}\vskip-7pt\hbox{$|\,/$}\vskip-7pt\hbox{o--o}}}  
 + 4 \;
\vcenter{\vbox{\hbox{o\phantom{--}o}\vskip-7pt\hbox{$|$\hskip10pt $|$}\vskip-7pt\hbox{o--o}}} + 7
\vcenter{\vbox{\hbox{\hskip6.5pt o--o}\vskip-7pt\hbox{\hskip5pt$\bigwedge$}\vskip-7pt\hbox{o--o}}} + 2\;
\vcenter{\vbox{\hbox{o--o}\vskip-7pt\hbox{$|$\hskip10pt $|$}\vskip-7pt\hbox{o--o}}} + 5\;
\vcenter{\vbox{\hbox{o--o}\vskip-7pt\hbox{$|\,/\,|$}\vskip-7pt\hbox{o--o}}} + 
\vcenter{\vbox{\hbox{o--o}\vskip-7pt\hbox{$|\,/\,\hskip-7pt\backslash\,|$}\vskip-7pt\hbox{o--o}}} 
\right) {t^4 \over 4!} + O(t^5).
\end{eqnarray}
This diagrammatic rule is similar to the Mayer theory
\cite{hansen,mayer-theory} for equilibrium fluid system.  The major
difference is the prefactor.  In Mayer theory, the integer factor
is simply the number of different ways of labeling a graph,
without imposing an ordering.  It turns out that the parallel is
deeper than this, see Table~\ref{tab:compare}.  Just like the
topological reduction in Mayer theory, by some transformation,
Given \cite{given} arrived at an expansion involving only the
star graphs (irreducible graphs).  Let $\phi = d\rho/dt$ be the
rate of adsorption, which is proportional to the area available
for deposition.  We have the following virial expansion
\begin{equation}
  \ln \phi = \sum_{n=1}^\infty b_n  { \rho^n \over n! },
\end{equation}
where $b_n$ is the sum of contributions from all $(n+1)$-point
star graphs with the RSA multiplicity factor.  The first few
terms are
\begin{equation}
\ln \phi  =  (\hbox{o--o}) \rho + 
(\vcenter{\vbox{\hbox{\hskip6.5pt o}\vskip-7pt\hbox{\hskip5pt$\bigwedge$}\vskip-7pt\hbox{o--o}}}
 ) { \rho^2 \over 2!} + \left( 2\;
\vcenter{\vbox{\hbox{o--o}\vskip-7pt\hbox{$|$\hskip10pt $|$}\vskip-7pt\hbox{o--o}}} + 5\;
\vcenter{\vbox{\hbox{o--o}\vskip-7pt\hbox{$|\,/\,|$}\vskip-7pt\hbox{o--o}}} + 
\vcenter{\vbox{\hbox{o--o}\vskip-7pt\hbox{$|\,/\,\hskip-7pt\backslash\,|$}\vskip -7pt\hbox{o--o}}} 
\right) {\rho^3 \over 3!} + O(\rho^4).
\end{equation}
The diagrammatic rules can also be applied to lattice models
when the Mayer function and integrals are properly interpreted.
In fact, Hoffman \cite{hoffman} has derived a similar expansion
for a lattice cooperative sequential adsorption model, while
Tarjus {\sl et al} \cite{tarjus-schaaf-talbot-JSP} obtained
diagrammtic expansion from a very different method.

\begin{table}[tb]
\begin{tabular}{|l|l|l|}
\hline\hline
expansion variable &  fugacity $z$  & time $t$ \\
\hline
quantities & pressure $ \displaystyle \beta P = { \ln \Xi \over V} $  
           & density $ \displaystyle \rho$  \\
           & $ \displaystyle \phi^{eq} = { \rho \over z} =  
                                  { 1\over V} {d \ln \Xi\over dz}$ 
           & adsorption rate $ \displaystyle \phi = { d\rho \over dt}$ \\
\hline
basic expansion & $ \displaystyle \beta P = { \ln \Xi \over V} = 
                             \sum_{n=1}^\infty a_n { z^n\over n!} $
               &  $ \displaystyle \rho = 
                               \sum_{n=1}^\infty a_n { t^n\over n!}$ \\
\vtop {\hsize=2cm \bigskip\noindent graphs}   
       & \vtop{\hsize = 5.25cm\baselineskip=11pt \raggedright\vskip0pt
                 $a_n = $ sum of all topologically distinct, 
           labeled, connected $n$-point graphs. \smallskip } 
         & \vtop{\hsize = 5.25cm\baselineskip=11pt \raggedright\vskip0pt
               $a_n =  $ sum of all topologically distinct, 
           labeled, connected $n$-point graphs, subject to the labeling 
           constraint that a vertex $i$ must be connected to a vertex with 
           label less than $i$.\medskip }  \\
\hline
virial expansion   
               & $ \displaystyle \ln \phi^{eq} = \ln {\rho \over z} = 
                                   \sum_{n=1}^\infty b_n { \rho^n\over n!}$ 
               & $ \displaystyle  \ln \phi = 
                                  \sum_{n=1}^\infty b_n { \rho^n\over n!}$ \\
\vtop{\hsize = 2.0cm \bigskip\noindent topological reduction}   & 
           \vtop {\hsize = 5.25cm\baselineskip=11pt\raggedright\vskip0pt
             $b_n = $ sum of all topologically distinct, 
           labeled, doubly connected (irreducible) $(n+1)$-point graphs.
           \smallskip
           \noindent $\displaystyle \beta P = 
            \rho - \sum_{n=1}^\infty { b_n \over n\!+\!1} 
            { \rho^{n+1}\over (n\!-\!1)! }.$
           } 
         & \vtop {\hsize = 5.25cm\baselineskip=11pt\raggedright\vskip0pt
            $b_n = $ sum of all topologically distinct, 
           labeled, doubly connected (irreducible) $(n+1)$-point graphs, 
           subject to the labeling 
           constraint that a vertex $i$ must be connected to a vertex with 
           label less than $i$.\medskip }  \\
\hline\hline
\end{tabular}
\caption[table compare]{Comparison of equilibrium Mayer expansion and 
random sequential adsorption.}
\label{tab:compare}
\end{table}

\begin{figure}[tb]
\center{\leavevmode\epsfxsize=0.5\hsize\epsfbox{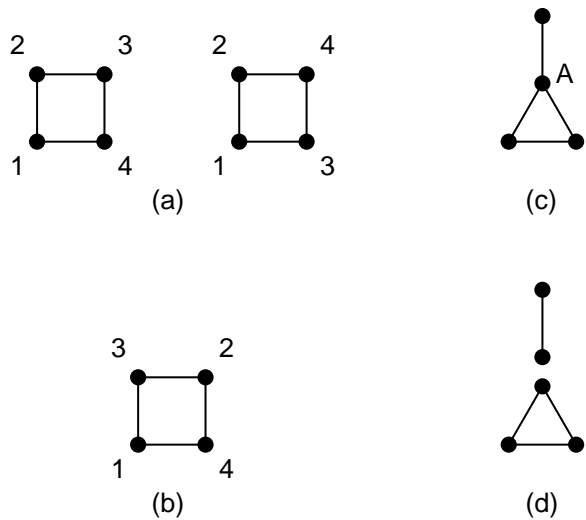}}
\caption[graphs]{Some concepts of graph theory: 
(a) The labeling of the vertices consistent with RSA constraints;
(b) this graph appears in equilibrium theory but not in RSA; (c)
a connected graph with an articulation point, labeled A; (d) the
connected graph (c) is decomposed into two star graphs
(irreducible graphs).}
\label{fig:graph}
\end{figure}

The computation of the cluster integrals can be further
simplified by noting that if a connected graph can be decomposed
into several star graphs, the value of the connected graph is
equal to the products of the values of star graphs.  Because of
this, we need only to evaluate the values of star integrals.  An
articulation point is a vertex belonging to the graph, such that
when this vertex and its connecting edges are removed, the graph
becomes disconnected.  A star graph does not have an
articulation point.  The decomposed graphs of a connected graph
are two or more graphs with the articulation point duplicated in
each of the separated graphs when the articulation point is
removed, with the edges still attached to the graphs.
Figure~\ref{fig:graph} demonstrates some of the concepts.

\begin{table}[tb] 
\begin{tabular}{|r|r|r|r|}
\hline\hline
n & hard disc & oriented square & nearest neighbor exclusion \\
\hline
0 & 1 & 1 & 1 \\ 
1 & 2 & 4 & 5 \\
2 & $4+3\sqrt{3}/\pi$ & 23 & 37 \\
3 & $8 + 14\sqrt{3}/\pi +44/\pi^2$ & 168 & 349 \\
4 & 86.02824 & $105895\over72$ & 3925\\
5 & & $6709687\over450$ & 50845 \\
6 & & $1385692277\over8100$ & 742165\\
7 & & $867234755659\over396900$ & 12017245\\
8 & & $7953008984179 \over 259200$ &213321717 \\
9 & & &4113044061 \\
10& & &85493084853 \\
11& & &1903886785277 \\
12& & &45187885535477 \\
13& & &1137973688508989 \\
14& & &30289520203949205 \\
15& & &849248887429012733 \\
16& & &25007259870924817749 \\
17& & &771322713104711008093 \\
18& & &24860884250598911650645 \\
19& & &835568036857675195155997 \\
20& & &29227671255970830546587445\\
\hline\hline
\end{tabular}
\caption[series]{The series expansion coefficient $S(n)$
for the hard discs \cite{dickman-wang-jensen}, 
oriented squares [this work], and nearest neighbor exclusion 
on square lattice \cite{gan-wang-JCP}.} 
\label{tab:cont}
\end{table}

In the work of Dickman {\sl et al} \cite{dickman-wang-jensen},
the series were obtained this way, using the existing cluster
integrations in the fluid Mayer expansion.  For the deposition
of discs, the series was obtained up to $S(4)$.  For the squares
of fixed orientation, the series is based on the result of
one-dimensional cluster integrals
\cite{hoover} which were already computed as early as in 1962 
up to order 6.  Note that for oriented hypercubes in $d$
dimensions, a cluster integral is equal to the one-dimensional
cluster integral raised to the power $d$.  This fact was used to
obtain two-dimensional results.  Bonnier {\sl et al}
\cite{bonnier-series} obtained results for $d = 2$, 3, and 4 
for RSA of oriented squares, cubes, and hypercubes, 
using an extrapolation procedure
from a lattice to continuum.  In
Table~\ref{tab:cont} we collect some of the series coefficients.

The computational complexity of the RSA on continuum is perhaps
more than $(n!)^3$.  One algorithm that we have implemented in this
work consists of two steps.  The first step reduces the labeled
graphs to unlabeled graphs, directly from defining
Eq.~(\ref{eq:Snf}).  This gives us the RSA topologically
distinct labeling factor.  This step is purely graph-theoretic,
and is independent of the detail of the problem.  There are
$\prod_{k=1}^n ( 2^k-1) = O(2^{n^2/2})$ labeled graphs at order
$n$.  To reduce the graphs to unlabeled graphs, we consider all
the $n$ factorial possible permutations of the labels, taking
the ``largest one'' as the canonical representation of the
(unlabeled) graph.  The largest one is the one with incidence
matrix viewed as a big binary integer giving biggest value.  
Since there are $O(
2^{n^2/2} )$ graphs, each needs $n!$ transformations, the
overall time complexity is $O( 2^{n^2/2} n!)$.  The above step
reduces the number of graphs by a factor of about $1/n!$.  Much
fast algorithms exist which generate each unlabeled graph
exactly once.  We used McKay's {\tt geng} program \cite{geng},
which greatly speeds up this part of the computation.

The second step is to compute the integrals.  computation of the
cluster integrals can be done exactly for the discs only for $n
\leq 3$.  For the one-dimensional segments, it can be done in
time of $O(n^4 n!)$.  The result raised to the $d$-th power
gives the $d$-dimensional hypercubic result.  For a given
cluster integral, the set of $f_{ij}$ gives a set of
constraints.  The integral is in fact equal to the volume of a
convex polygon in an $n$-dimensional space.  This
$n$-dimensional space is partitioned into $n!$ regions,
characterized by $ x_{i_1} < x_{i_2} < \cdots x_{i_n}$ where
$(i_1,i_2, \cdots, i_n)$ is a permutation of $(1,2, \cdots, n)$.
Once this extra constraint is given, the integration limits can
be written down explicitly
\cite{hoover}. The integrals in each sector can be evaluated
symbolically in polynomial time. 

This lastest program gives only two extra new terms after 35
hours of CPU time on a 600MHz Alphastation.  The Dickman {\sl et
al} \cite{dickman-wang-jensen} and Bonnier {\sl et al}
\cite{bonnier-series} results were reproduced in 5 seconds.  To
obtain the next term requires about $10^3$ times more in
computer time, which is not possible within availability of our
computer resources.

\subsection{Other formulations of series expansions}

Beside the rate equation approach described in the previous
three subsections, there are a number of other methods to obtain
the series.  Baram and Kutasov
\cite{baram-kutasov} and Baram and Fixman \cite{baram-fixman}
used Ising spin notation and master equation to relate expansion
coefficients to a counting problem of lattice animals as
follows.  The particle density is expanded as
\begin{equation}
  \rho(t) = \sum_{n=1}^\infty (-1)^{n+1} a_n { (1 - e^{-t})^n \over n!}.
\label{eq:animal}
\end{equation}
For the nearest-neighbor exclusion model, the coefficient $a_n$
is the number of connected lattice animals with $n$ points that
can grow from a single point.  Unlike the rules embedded in the
Eq.~(\ref{eq:Sn}) for $S(n)$ the new lattice animals are
generated from perimeter sites only.

We can get this result from the rate equation point of view as
follows.  Let $P(W)$ be the probability of a set of connected sites
$W$ for each of which a particle can be put in for sure. A
large set $G$ must be known to be empty.  However, the
correspondence between $W$ and $G$ is many to one.  The rate
equation for $P(W)$ is then
\begin{equation}
 { d P(W) \over dt } = - m P(W) - 
 \sum_{x} P(W \cup \{x\}),
\end{equation}
where $m=|W|$ is the number of sites in $W$, and $x$ runs over
the perimeter sites of $W$ (for the nearest neighbor exclusion
model). Let us transform $t$ to $y = 1 - e^{-t}$, and $Q(W,y) =
(1-y)^{-m} P(W,t)$, then the new equation is
\begin{equation}
 { d Q(W) \over dy} = - \sum_{x, {\rm\ perimeter\ of\ } W} 
Q(W \cup \{x\}).
\end{equation}
This naturally gives Eq.~(\ref{eq:animal}) and rule for $a_n$.

Dickman {\sl et al} \cite{dickman-wang-jensen} gave an operator
formulation of the master equation.  The formal solution of the
master equation is used to derive expansion rules identical to
the rate equation approach.  For the continuum problem, Schaaf
and Talbot
\cite{schaaf-talbot} gave small density expansions based
on results in scaled particle theory
\cite{reiss-frisch-lebowitz}.  The final result is equivalent to
the time expansion, but the procedure is quite different.
Consider the average fraction $\phi$ of surface available to the
center of a new disc, $\phi = d \rho/dt$.  One has
\begin{equation}
  \phi = 1 - S_1 + S_2 - S_3 + S_4 - \cdots,
\end{equation}
with 
\begin{equation}
S_n = {1 \over n!} \int\int \cdots \int dx_1 dx_2 \cdots dx_n 
\> A_n(x_1, \cdots, x_n)\> \rho^{(n)}(x_1,\cdots, x_n),
\end{equation}
where $A_n(x_1,\cdots,x_n)$ is the area common to the exclusion
circle of $n$ particles adsorbed at the positions defined by the
vectors $x_1$, $x_2$, \dots, $x_n$; i.e., it is the area of the
set $D(x_1) \cap D(x_2) \cdots \cap D(x_n)$.  The
multi-variable function $\rho^{(n)}(x_1, \cdots, x_2)$ is the
general $n$-particle distribution function, which is
proportional to the probability that a particle will be found in
the elements of surface $dx_1$ at $x_1$, a second in $dx_2$ at
$x_2$, {\sl etc}.  Schaaf {\sl et al} observed that the leading
order in a density expansion for $S_n$ is equal to a
corresponding equilibrium system.  This allows them to find
exact expression for $S_n$ correct to third order in density
\cite{schaaf-talbot,schaaf-talbot-JCP,talbot-schaaf-tarjus-MP,ricci-etal}.
Tarjus {\sl et al} \cite{tarjus-schaaf-talbot-JSP} derived
Kirkwood-Salsburg-like equations for $\rho^{(n)}(x_1,
\cdots,x_n)$.  From these equations, a diagrammatic expansion in
density is obtained.  Some what opposite in spirit to the rate
equation approach, Given \cite{given} formulates nonequilibrium
problems as quenched, multi-species equilibrium problems.  He also
gave the topological reduction in graph expansion, showed clearly
the parallel between RSA and equilibrium system.  Caser and
Hilhorst \cite{caser-hilhorst} presented series expansion directly for
the jamming coverage.  The approach is interesting, but the
method is not very accurate.

\begin{figure}[tb]
\center{\leavevmode\epsfxsize=0.8\hsize\epsfbox{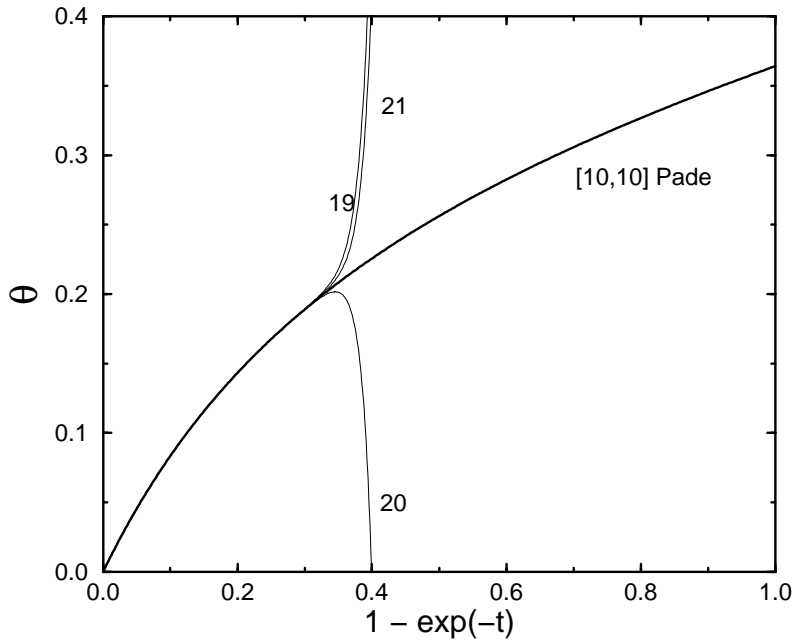}}
\caption[series and Pade]{Series results for the RSA coverage on
a square lattice of the monomer with nearest neighbor exclusion
model.  The curves labeled 19, 20, and 21 are the direct time
series truncated to 19, 20, and 21 orders, respectively.  The
line labeled ``[10,10] Pade'' is $N=D=10$ Pad\'e approximant in
variable $y$, defined by Eq.~(\ref{eq:y1}), with $ b = 1.05$.
For convenience of viewing the whole range of $t$, we plot the
curves with variable $1 - \exp(-t)$.}
\label{fig:pade}
\end{figure}

\subsection{Analysis of the series}
The series obtained for the coverage in $t$ has a finite radius
of convergence, see Fig.~\ref{fig:pade}.  The fundamental
question is whether we can give good approximation valid for the
whole time domain $t \in [0, \infty)$.  Although we do not have
rigorous proof, the numerical procedure which we shall discuss
below indicates a positive answer.

The standard technique of extending the radius of convergence is
the Pad\'e approximation method
\cite{guttmann,baker}.  
Given a series $f(x)$ to order $L$, we determine two polynomials
$P_N(x)$ and $Q_D(x)$ of degree $N$ and $D$ respectively, such
that
\begin{equation}
f(x) - { P_N(x)\over Q_D(x) } = O(x^{L+1}), \qquad N+D\leq L. 
\label{eq:pade}
\end{equation}
It is a powerful way of extending the domain of convergence of
the original series.  The polynomials $P_N(x)$ and $Q_D(x)$ with
$Q_D(x) = 1 + b_1 x + \cdots$ is usually determined uniquely
for the given $f(x)$.

To accelerate the convergence further, new variables are
introduced, e.g.,
\begin{equation}
y = 1 - \exp\Bigl(-b ( 1 - e^{-t})\Bigr).
\label{eq:y1}
\end{equation}
The functional form is chosen in such a way so that the series
in the new variable $y$ most closely resembles the asymptotic
behavior of the coverage $\theta(t)$ at large $t$.  For small
$t$, $y$ is proportional to $t$.  The above specific form is
encouraged by the exact solution of the one-dimensional dimer
problem.  In fact, we have $\theta = y$ with $b=2$ in such case
(see Eq.~(\ref{eq:1dexact})).  The convergence among various
Pad\'e approximants is improved greatly by the transformation.
This form of transformation works well for all lattice models
that we have studied.  For continuum models, Dickman {\sl et al}
\cite{dickman-wang-jensen} considered
\begin{equation}
 y = 1 - { 1 \over \sqrt{ 1 + b t} } 
\label{eq:y2}
\end{equation}
for the RSA of discs in two dimensions and 
\begin{equation}
 y = 1  -  { 1 + \ln[ 1 + (b-1) t] \over 1 + b t}
\label{eq:y3}
\end{equation}
for the oriented squares, which match the asymptotic law of
$t^{-1/2}$ (known as Feder's law \cite{feder}) and $ \log t/t$
law (Swendsen \cite{swendsen}), for discs and oriented squares,
respectively.

Other methods of analysis were also used.  For example, Fan and
Percus \cite{fan-percus} used exact results of solvable systems
as reference systems which in some sense are close to the actual
system to speed up the convergence.  The coverage of the nearest
neighbor exclusion model on a square cactus fractal lattice is
\cite{fan-percus}
\begin{equation}
  \theta(z(t)) = {1 \over 6}  + {1 \over 3} z -
{1\over 6} (1-z)^4,
\end{equation}
where variable $z$ is related to the original variable $t$ by
\begin{equation}
  1 - e^{-t} = 3 \int_0^z { dz' \over 1 + 2(1-z')^3  } .
\label{eq:percus}
\end{equation}
We note that the coverage $\theta$ is a truncated power series
in $z$ for the cactus system.  The idea is that if we use the
same variable $z$, relating $t$ through Eq.~(\ref{eq:percus}),
the nearest neighbor exclusion model on square lattice should
have a series in $z$ with much improved convergence.

Baram {\sl et al}  \cite{baram-kutasov,baram-fixman} used
Euler transform 
\begin{equation}
 y = { 1 - u \over c + 1 - u}, \quad u = e^{-t},  \quad c \approx 1,
\end{equation} 
and Levin convergence acceleration method \cite{levin} for the
series analysis.  It appears that the quality of the results is
comparable to other methods.

The algebraic manipulations can be done with great ease by
symbolic packages such as Maple or Mathematica.  Mathematica
contains function to find the Pad\'e approximation of a
series.  The results of the analysis can then be plotted
within the software.  Here is an example of simple Mathematica
code to form the series, do the transformation, call the Pad\'e
function, and finally evaluate the jamming coverage.  The {\tt
SetPrecision[1.35,40]} command makes the subsequent computation in
40-digit precision.

{\baselineskip=10pt
\begin{verbatim}
<<Calculus`Pade`
maxord=17;
P[0] = 1;
P[1] = -4;
P[2] = 28;
P[3] = -268;
 ... (omitted)
P[17] = -6058617368871081964076;
theta = 1 - Sum[P[i]*t^i/i!, {i,0,maxord}] + O[t]^(maxord+1);
b = SetPrecision[1.35,40];
f = theta /. t -> - Log[1 + Log[1-y]/b] + O[y]^(maxord+2);
yinf = 1 - Exp[-b];
pd = Pade[f, {y, 0, 8, 8}]; 
thetainf = pd /. y -> yinf
\end{verbatim}
}
The Pad\'e approximant for the coverage of
the dimer on square lattice is given in the {\tt pd} variable as
\begin{eqnarray}
&&  \Big(2.962963\,y + 0.03206897\,{y^2} - 2.195246\,{y^3} 
 - 1.073721\,{y^4}  +  0.9207869\,{y^5} + 0.5556586\,{y^6}  \nonumber \\
&& - 0.04386743\,{y^7} - 0.05303456\,{y^8} \Big)\Big/\Big( 
   1 + 1.733045\,y - 0.2568919\,{y^2} - 1.942572\,{y^3}   \nonumber  \\
&&  - 0.5852424\,{y^4} + 0.7908992\,{y^5} + 0.4421557\,{y^6} - 
     0.0493306\,{y^7} - 0.0513337\,{y^8} \Big),  
\end{eqnarray}
where $y$ is given by Eq.~(\ref{eq:y1}) with $b = 1.35$.  The
result is stable against variation in $b$.  Error can be
estimated from the convergence of various Pad\'e approximants.
The above Pad\'e approximant is accurate to $10^{-5}$ for all
$t\geq 0$.  Such a high degree of accuracy embedded in a simple
formula is remarkable.

\begin{figure}[tb]
\center{\leavevmode\epsfxsize=0.7\hsize\epsfbox{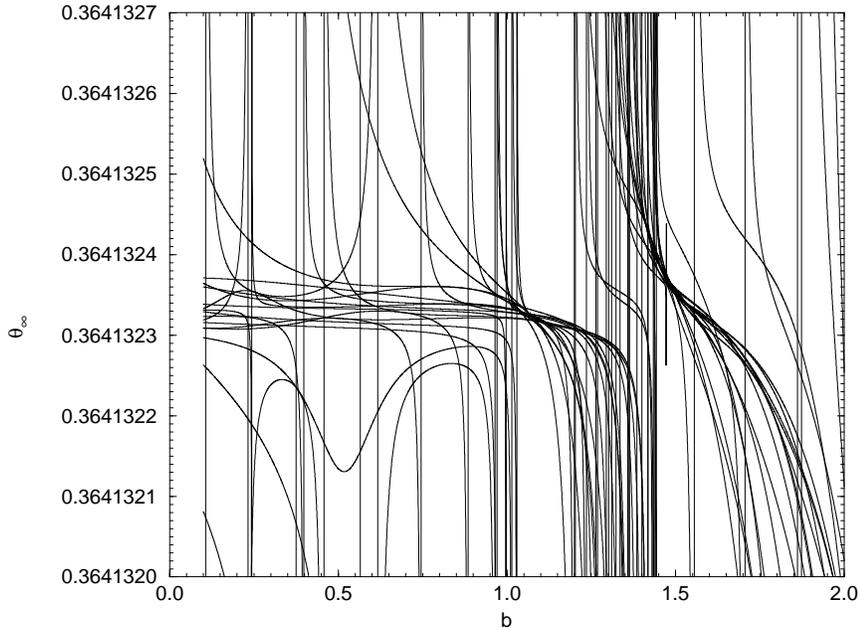}}
\caption[nn exclusion RSA on square lattice]{Pad\'e approximant 
estimates for the jamming coverage $\theta(\infty)$ as a function
of the transformation parameter $b$, for the monomer RSA with
nearest neighbor exclusion on a square lattice.  The
transformation is given by Eq.~(\ref{eq:y1}).  (from
ref.~\cite{gan-wang-JCP}).}
\label{fig:sqrnn21}
\end{figure}

\begin{figure}[tb]
\center{\leavevmode\epsfxsize=0.7\hsize\epsfbox{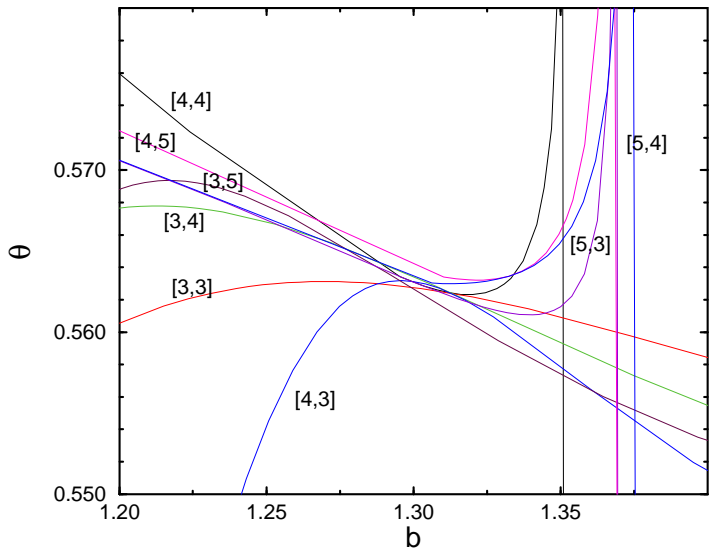}}
\caption[Oriented Square Pade]{Pad\'e approximant 
estimates for the jamming coverage $\theta(\infty)$ as a function
of the transformation parameter $b$, for the oriented square.
The number $[N,D]$ is the order of the Pad\'e approximant. 
The transformation is given by Eq.~(\ref{eq:y3}). }
\label{fig:pdSQ}
\end{figure}

We set $b$ to be a free variable so that when different orders
of Pad\'e approximants are applied to $\theta(y)$, a crossing
region between different orders of Pad\'e approximants is to be
located so as to give a good estimate of jamming coverage
$\theta(\infty)$.  Some of the new series results and analyses
are presented in Gan and Wang \cite{gan-wang-JCP}.  As an
example, Fig.~\ref{fig:sqrnn21} shows the crossing region of
Pad\'e approximants of orders $[N,D]$ with $19 \leq N + D \leq
21$, $D
\ge 8$, $N \ge 8$, for the series of monomer RSA with nearest
neighbor exclusion on a square lattice, giving an estimate of
$\theta(\infty) = 0.3641323(1)$, where the last digit in
parentheses denotes the uncertainty in the preceding digit.
This estimate is in good agreement with the estimate of
$\theta(\infty) = 0.3641330(5)$ by Baram and Fixman
\cite{baram-fixman}. Analyzing the series 
using the square cactus as a reference model
\cite{fan-percus}, through the transformation Eq.~(\ref{eq:percus}),
we obtain $\theta(\infty) = 0.364132(1)$, less accurate than that
using transformation Eq.~(\ref{eq:y1}). All these estimates from the
series analysis agree well with the simulation result of
$\theta(\infty) = 0.36413(1)$ \cite{meakin-cardy-etal}.

The estimates for the jamming coverage on
continuum systems are much less accurate, due to limited number of terms
available in the series. In Fig.~\ref{fig:pdSQ} we present the Pad\'e
estimates for the oriented squares as functions of the parameter
$b$.  Best convergence occurs around $b=1.3$, given
$\theta(\infty) = 0.5623(4)$.  The additional new terms do not
improve the result of Dickman {\sl et al}
\cite{dickman-wang-jensen} very much.

\begin{table}
\begin{tabular}{|l|r|l|l|}
\hline\hline
\hfil {\em Model} & \multicolumn{2}{c|}{\em Series } & {\em Monte Carlo } \\
\cline{2-3}
  &  {\em Order $n$} & { \em Pad\'e $\theta(\infty)$ } & $\theta(\infty)$\hfill \\  
\hline
NN (square lattice) &   21  & 
\hbox to 3.6 cm { 0.3641323(1) \hfill \cite{gan-wang-JCP}}  &  
\hbox to 3.6 cm { 0.36413(1) \hfill \cite{meakin-cardy-etal}} \\
dimer (square lattice) & 18 & 
\hbox to 3.6 cm { 0.906823(2) \hfill \cite{gan-wang-JCP}} &
\hbox to 3.6 cm { 0.906820(2) \hfill \cite{wang-pandey}} \\
NN (honeycomb) & 24 & 
\hbox to 3.6 cm { 0.37913944(1) \hfill \cite{gan-wang-JCP}} & 
\hbox to 3.6 cm { 0.38(1) \hfill \cite{widom}} \\
dimer (honeycomb) & 22 & 
\hbox to 3.6 cm { 0.8789329(1) \hfill \cite{gan-wang-JCP}} & 
\hbox to 3.6 cm { 0.87889 \hfill \cite{nord-evans-JCP85}}  \\
NNN (square lattice) &  14 & 
\hbox to 3.6 cm { 0.186985(2) \hfill \cite{baram-fixman}} & 
\hbox to 3.6 cm { 0.186983(3) \hfill \cite{privman-wang-nielaba}} \\
\hline
hard discs & 5 & 
\hbox to 3.6 cm { 0.5479 \hfill \hfill \cite{dickman-wang-jensen}} & 
\hbox to 3.6 cm { 0.5470690(7) \hfill \cite{wang-ijmp}} \\
oriented squares & 9 &  
\hbox to 3.6 cm { 0.5623(4) \hfill [this work]} & 
\hbox to 3.6 cm { 0.562009(4) \hfill \cite{brosilow-ziff-vigil}} \\
\hline\hline
\end{tabular}
\caption[table, Jamming]{Some of the most accurate jamming coverages by 
series analysis and 
Monte Carlo computer simulation.  NN stands for nearest neighbor exclusion,
NNN for nearest and next neighbor exclusion.  The
order $n$ is the highest known term in the time expansion of
$\theta(t)$. The numbers in square brackets are references. }
\label{tab:jamming}
\end{table}

In Table~\ref{tab:jamming}, we collect the results of jamming
coverage compared with Monte Carlo results.  It is clear, if the
jamming coverage can be estimated with great accuracy, so is the
whole function $\theta(t)$.  The results for continuum models
are less accurate.
Nevertheless, we still obtain quite reasonable estimates.

\section{Monte Carlo simulation}

A large variety of RSA models have been simulated.  Lattice
models are easy to simulate with good accuracy.  Random dimer
filling is given by Nord and Evans \cite{nord-evans-JCP85} and
Wang and Pandey \cite{wang-pandey}.  Nearest neighbor exclusion
models are simulated by Meakin {\sl et al}
\cite{meakin-cardy-etal}.  Square blocks of $n\times n$ are
simulated by Nakamura \cite{nakamura}, and by Barker and Grimson
\cite{barker-grimson}.  
Square blocks with an emphasis of approaching to the
continuum limit are studied by Privman {\sl et al} 
\cite{privman-wang-nielaba}. 

On continuum, hard-disc system is simulated by Akeda and
Hori \cite{akeda-hori}, Finegold and Donnell
\cite{finegold-donnell}, Tanemura \cite{tanemura},
Feder \cite{feder}, Hinrichsen {\sl et al}
\cite{hinrichsen-etal}, Meakin and Jullien \cite{meakin-jullien},
and Wang \cite{wang-ijmp}.  The oriented squares were of interest
\cite{akeda-hori} because of Pal\'asti conjecture 
\cite{palasti}, which says that the jamming coverage for the
two-dimensional oriented square is equal to the 
square of the one-dimensional line-segment jamming coverage.  
This conjecture turns out to be false
\cite{brosilow-ziff-vigil}. 

Other more elaborated geometries are considered such as randomly
oriented ellipses \cite{talbot-tarjus-schaaf-PRA} and lines
\cite{sherwood,ziff-vigil},
randomly oriented rectangles \cite{vigil-ziff},
and other anisotropic particles \cite{viot-etal},
spheroids \cite{adamczyk},
mixtures of different sizes 
\cite{meakin-jullien,talbot-schaaf-mixture,tarjus-talbot-mixture},
The interests are the jamming coverage and the asymptotic
laws for large time.  Clearly, these complicated
problems are difficult for systematic series expansions
\cite{ricci-etal}.  
Lattice models with dimers on square lattice deposited with
different probabilities in two orientations
\cite{oliveira-tome-dickman}, with line segments   
\cite{manna}, mixtures 
\cite{svrakic-henkel,bonnier-mixture,sinkovits-pandey},
and random walks \cite{wang-pandey,budinski} are also studied.

We mention here briefly two related generalizations of standard
RSA. The first of these is multilayer RSA
\cite{nielaba-privman-wang}.  The jamming densities 
as a function of the height show interesting power-law behavior
\cite{hilfer-wang}.  The second generalization is RSA with
diffusional relaxation 
\cite{gan-wang-PRE,privman-nielaba-europhys,privman-barma,wang-defect,pereira,bonnier,eisenberg-baram}.
When diffusional relaxations are introduced, 
many of the intrinsic properties of RSA changes.
Most of these generalized models are studied by Monte Carlo
computer simulation.

\subsection{Simple algorithms}

Monte Carlo simulation method has the advantage of very little
programming investment, quick and robust results.  Random
sequential adsorption, especially lattice models, can be
simulated on a computer rather easily.  The basic variables
(declared as an array) are the occupations of the lattice sites.
At each time step, a site $x$ is picked at random, if some
neighborhood $D(x)$ is empty, we put down the particle by reset
the array values in these entries of involved sites.  Note that
one is simulating a time-discretized version of the RSA model,
while the rate equation approach considers continuous time.  The
discretization step is $1/L^d$ where $L$ is the linear dimension
of the system.  However, discretization effect is unimportant
since we can simulate very large systems easily.

With continuum models, since we can not have occupation
variables like the lattice models, more tricks are needed for
efficiency.  The coordinates of the particles are stored to
register the locations.  To speed up the checking for
overlapping, ``coarse-grained occupation variables'' are
needed to indicate where are the particles in a given range, a
technique borrowed from molecular dynamics simulation.  For
example, in RSA of discs, the total $L\times L$ surface area is
partitioned into regions of squares of size $1 \times 1$.  A
disc at $(x,y)$ belongs to the unit square at $(\lfloor x
\rfloor,\, \lfloor y \rfloor)$, where the notation 
$\lfloor x \rfloor$ represents the floor or integer part of $x$.
An array of linked lists of disc coordinates is used to
represent the deposited discs.  With this data structure, it is
suffice just to check the discs located at the center and eight
surround squares for overlaps.

The above simple methods have a severe shortcoming that the
study of late-stage process is very time-consuming.  When the
jamming configuration is about to be reached, much of the area
is blocked.  Only tiny disconnected regions are available for
further deposition.  Since we choose the sites at random with a
uniform probability distribution, the available sites are hard
to find.  More advanced methods cleverly discover these regions
and suggest deposition only in regions which are potentially
possible for adsorption.

\subsection{Event driven algorithms}

First we consider the lattice models. On a lattice, since there
are only finite number (of order $N=L^d$) of possibilities for
deposition, each with equal probability in the original
definition of the model, we can classify the deposition
possibilities as possible and impossible attempts.  For example,
for the nearest neighbor exclusion model, each site is
associated with a deposition attempt.  The possible sites are
those with center and four neighbors empty; the impossible sites
are those with any one of the five sites occupied.  Let us
assume that we have $N_a$ possibilities for the available type
and $N_i$ for the impossible type. Clearly, $N_a + N_i = N$.  In
the normal simulation, we hit the first available type with
probability $p = N_a/N$ and other type with $1-p=N_i/N$.
Consider a total of $k$ attempts such that the first $k-1$
failed to deposit a particle because one hits already blocked
sites and the $k$-th attempt is successful and is deposited.
What is the probability that such an event occurs?  Each failed
attempt occurs with probability $(1-p)$, so $k-1$ consecutive
fails have probability $(1-p)^{k-1}$.  Since each failed attempt
does not change the current state of the system, the next
attempt is totally independent of the previous attempts.  The
successful attempt has probability $p$ so the total probability
is
\begin{equation}
   P_k = p (1-p)^{k-1}, \quad k = 1, 2, \cdots. 
\label{eq:event}
\end{equation}
Imagine that we do not actually do the $k-1$ failed attempts,
and pick one at random from the available sites and deposit
the particle.  From the above discussion, this is equivalent to
$k$ attempts in units of the original time scale.  However, $k$
is not a definite number, rather it follows the probability
distribution, Eq.~(\ref{eq:event}).

Event-driven type of algorithms was invented long-time ago for
the simulation of Ising models
\cite{bortz-kalos-lebowitz}, which is known as N-fold way.
An event-driven RSA algorithm
\cite{wang-ijmp,brosilow-ziff-vigil}
on lattice is as follows: pick deposition site from the list of
available sites at random.  Increase the time $t$ by $k/L^d$
($L^d$ attempts are usually defined as unit time).  Update the
configuration and the list.

The random integer $k$ can be generated by transformation method
as follows:
\begin{equation}
k = 1 + \left\lfloor { \ln \xi \over \ln (1-p) } \right\rfloor,
\label{eq:k}
\end{equation}
where $\xi$ is a uniformly distributed random number between 0
and 1.  Some programming efforts are needed to update the list
efficiently.  After one deposition, not only the current site
but also some other sites in the neighborhood of the current
site are no longer available.  This should be correctly
reflected in the list.  The deposition also changes the value
$p$.

Perhaps the best application of the technique to lattice model
is the RSA of polymer chains \cite{wang-pandey}.  Due to a large
number of conformational states of polymer chains, the
deposition becomes very slow in the late stage.  An event-driven
method is, therefore, essential to obtain long-time results.
This is accomplished by identifying the early part of growing
chains, i.e., the partial chains and then classifying them
periodically as ``available'' or ``not available'' for the
deposition in an iterative fashion.  The subsequent depositions
are then made starting from the available partial chains
choosing at random.  The list of the partial chains is limited
by the available computer memory and governs the speed of the
program.  It appears that there is an optimal list size for a
given length of a chain.

Highly accurate Monte Carlo simulation results were obtained for
oriented squares \cite{brosilow-ziff-vigil} and for discs
\cite{wang-ijmp} on two-dimensional continuum.  In the RSA
simulation of discs, we incorporated two important ideas: (1)
divide the surface into small squares and make deposition
attempts only on squares that are not completely blocked; (2)
systematically reduce the sizes of the small squares after some
number of attempts and re-evaluate the availability of the
squares.

In this more elaborate method, we make attempts only on squares
which are potentially possible for depositions.  Thus the core
part is an algorithm which identifies correctly whether a square
is available for deposition.  Since the diameter of the discs is
$\sigma=1$, each disc excludes an area of circle of radius
$\sigma$ for further deposition.  If some area is completely
covered by a union of the exclusion zones of discs, that area is
not available for deposition.  We begin by classifying all the
squares of $a\times a$ with $a=1$ as available or unavailable
squares.  The available squares are put on a list.  Deposition
trials are taken only on the squares in the list, i.e., a square
in the list is chosen at random, deposition is attempted with a
uniform probability over the square. After certain number of
trials (we used $5 \times 10^5$), we rebuild the list, now with
squares of size $(a/2) \times (a/2)$, checking only those
squares on the old list.  Each old square is subdivided into 4
smaller squares.  This shrinking of the basic squares helps to
locate even extremely small available regions.  They will be hit
with very small probability if the area is always $1 \times 1$.
Because of integer overflow, the size of the squares is not
allowed to go arbitrarily small, but the shrinking is stopped at
$ a = 2^{-15}$.  Thus the smallest square has an area of
$2^{-30}
\approx 10^{-9}$.

For the standard RSA algorithm, each trial increases the time
$t$ by $1/A$, where $A = L^2$ is the area of the total surface.
In the even-driven algorithm, where only the potentially
successful depositions are tried, the clock must tick faster in
order to compensate for not making attempts in the completely
unsuccessful area.  The time increment for each trial on the
available squares is given by $k/A$ with $k$ generated using
Eq.~(\ref{eq:k}), with $p$ being the ratio of the area on which
we try our deposition to the total area; it is equal to
$a^2/L^2$ times the number of available squares.

An important piece of code of our RSA simulation program is the
identification of the squares which are fully covered by the
excluded area of discs.  For these squares we are sure that
depositions on them will be unsuccessful, and thus they will not
be on the list of candidates for trials.

\begin{figure}[tb]
\center{\leavevmode\epsfxsize=0.8\hsize\epsfbox{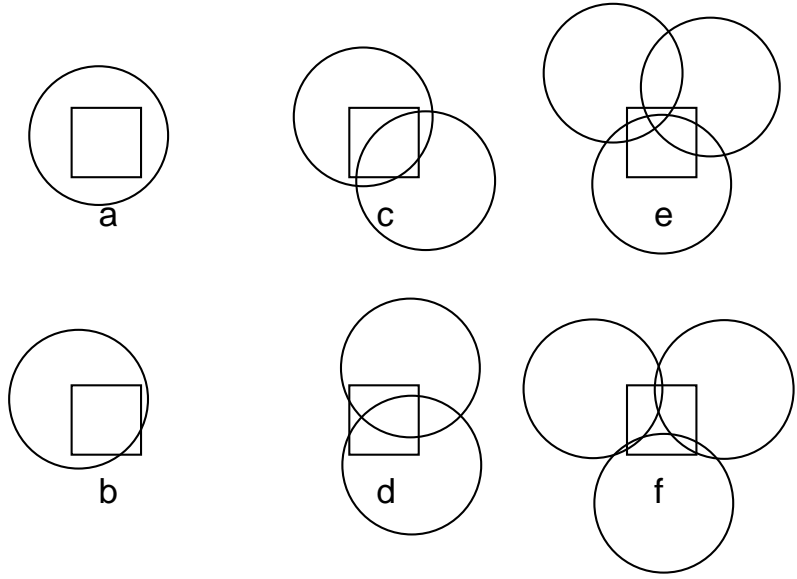}}
\caption{The six different situations for discs covering or
not covering a square.}
\label{fig:cover}
\end{figure}

The algorithm that we have devised is as follows.  Written as a
C programming language function, it returns a nonzero value if
the square is fully covered and a value 0 otherwise.  Part (1)
to~(5) is executed in the order given.

\begin{enumerate}

\item Find relevant discs to the current square.  A disc is
relevant if the square overlaps with the excluded region of the
disc.  Note that a disc has a diameter $\sigma$ and its excluded
region is a concentric circle of radius $\sigma$.

\item If all of the four vertices of the square are inside a 
single-disc excluded region, then it is already fully covered
(Fig.~\ref{fig:cover}a).  Function returns with value~1.

\item For a full coverage, each vertex of the square must
be at least in one of the excluded region of the relevant discs.
If any one of the vertices is not covered at all by any disc,
then the square is not fully covered (Fig.~\ref{fig:cover}b).
Function returns with value~0.

\item At this point, at least two discs are involved if the
function is not returned.  Mapping the edges of the square to a
one-dimensional line between 0 and 4, we find out all the line
segments which are covered by the excluded region of discs.  If
the four edges of the square are completely covered by the
discs, and we have exactly two discs, then the square is fully
covered (Fig.~\ref{fig:cover}c), function returns with value~2.
If there are segments not covered by discs, then we know that
the square is not fully covered (Fig.~\ref{fig:cover}d).
Function returns with value~0.

\item If the function is not returned, we know that there
are at least three discs.  Even if all the edges are covered,
the square can still contain uncovered area if three or more
discs are involved.  Three or more discs can form holes.  To
make the last check, the coordinates of the intersection points
of circles (with radius $\sigma$) are calculated.  In order for
the interior of the square to be covered, all the intersection
points of any pair of discs which lie inside or on the square
edges must be covered by a third disc.  Return with the number
of relevant discs if the above condition is satisfied
(Fig.~\ref{fig:cover}e); return value~0 if not
(Fig.~\ref{fig:cover}f).

\end{enumerate}

The order of various checks is arranged in such a way so that
the most frequent and easy situations are checked first. Even
though the worse case computational complexity goes as $n^2$,
where $n$ is the number of relevant discs, a definite conclusion
can be made typically much earlier.  Using this algorithm, Wang
\cite{wang-ijmp} obtained accurate jamming coverage
$\theta(\infty) = 0.547 069 0 (7) $ and confirmed to a high
accuracy the Feder's asymptotic law $\theta(t) \approx 
\theta(\infty) - c\, t^{-1/2}$.

\section{Conclusion}

We introduced the method of series expansions for both lattice
models and continuum models.  The computation of the series for
the lattice model can be reduced to a counting problem.  An
efficient implementation requires full use of the symmetry of
the problem.  For continuum models, systematic expansion in
terms of a class of graphs are available.  Both the problem of
reducing labeled graphs to unlabeled graphs and of computing
the cluster integrals are computationally difficult problems.
Thus, the series for continuum systems are rather short.  The
series can be analyzed through variable transformation and
Pad\'e approximation.  Monte Carlo simulation method is
versatile, and implementation is straightforward.  However,
efficient algorithms are available with more sophisticated data
structures.  The challenge to the series expansion and efficient
Monte Carlo simulation for RSA is to apply to more complicated
models which may describe better real experiment situations.

\section*{Acknowledgements}

The author thanks R. Dickman, C.~K. Gan, R. Hilfer, P. Nielaba,
and V. Privman.  Many of the original results presented in this
article were obtained together with them in fruitful
collaborations.  He also thanks E.~C. Chang for discussion on
graph algorithms.


\newpage
\begin{thebibliography}{01}

\bibitem{flory} P. J. Flory,  J. Am. Chem. Soc. 61 (1939) 1518.

\bibitem{renyi} A. Renyi, Publ. Math. Inst. Hung. Acad. Sci,
                3 (1958) 109; Selected Transl. Math. Stat. Prob.
                4 (1963) 203.

\bibitem{feder} J. Feder, J. Theor. Biol. 87 (1980) 237.

\bibitem{finegold-donnell} L. Finegold and J. T. Donnell,
               Nature, 278 (1979) 443.

\bibitem{onoda-liniger} G. Y. Onoda and E. G. Liniger,
               Phys. Rev. A 33 (1986) 715.

\bibitem{privman-wang-nielaba-PRB} V. Privman, J.-S. Wang, and
                 P. Nielaba, Phys. Rev. B. 43 (1991) 3366. 

\bibitem{pomeau} Y. Pomeau, J. Phys. A: Math. Gen. 13 (1980) L193.

\bibitem{swendsen} R. H. Swendsen, Phys. Rev. A, 24 (1981) 504. 

\bibitem{viot-tarjus-europhys-lett} P. Viot and G. Tarjus, 
                         Europhys. Lett. 13 (1990) 295.

\bibitem{tarjus-viot-PRL} G. Tarjus and P. Viot, Phys. Rev. Lett.
                           67 (1991) 1875. 

\bibitem{evans-review} J. W. Evans, Rev. Mod. Phys. 65 (1993) 1281.

\bibitem{bartelt-privman-review} M. C. Bartelt and V. Privman,
                  Int. J. Mod. Phys. B 5 (1991) 2883. 

\bibitem{ramsden}  J. J. Ramsden, J. Stat. Phys. 73 (1993) 853.

\bibitem{page} E. S. Page, J. R. Stat. Soc, B 21 (1959) 364.

\bibitem{mackenzie} J. K. Mackenzie, J. Chem. Phys. 37 (1962) 723.

\bibitem{cohen-reiss} E. R. Cohen and H. Reiss,
                      J. Chem. Phys. 38 (1963) 680.

\bibitem{gonzalez-hemmer} J. J. Gonzalez, P. C. Hemmer, and
                     J. S. H\o ye, Chem. Phys. 3 (1974) 228.

\bibitem{bonnier-boyer} B. Bonnier, D. Boyer, and P. Viot,
                        J. Phys. A: Math. Gen. 27 (1994) 3671.

\bibitem{evans-tree} J. W. Evans, J. Math. Phys. 25 (1984) 2527.

\bibitem{baram-kutasov-1d} A. Baram and D. Kutasov,
                        J. Phys. A: Math. Gen. 25 (1992) L493;
                        J. Phys. A: Math. Gen. 27 (1994) 3683.

\bibitem{fan-percus-ladder} Y. Fan and J. K. Percus, 
                            J. Stat. Phys. 66 (1992) 263.
                        
\bibitem{guttmann} A. J. Guttmann, in ``Phase Transitions and
              Critical Phenomena,'' Vol. 13. edited by C. Domb and 
             J. L. Lebowitz, Academic, New York, 1989.

\bibitem{domb} C. Domb, ``The Critical Point,'' Taylor \& Francis, 
                        London, 1996.

\bibitem{baker} G. A. Baker and P. Graves-Morris, 
                ``Pad\'e Approximants, Encyclopedia of 
                  Mathematics and its Applications, Vols. 13 and 14.
                  Addison-Wesley, Reading, 1981.
               G. A. Baker, ``Quantitative Theory of Critical Phenomena,''
               Academic Press, Boston, 1990.

\bibitem{evans-nord-JSP85} J. W. Evans and R. S. Nord, 
                  J. Stat. Phys. 38 (1985) 681.

\bibitem{nord-evans-JCP85} R. S. Nord and J. W. Evans,
                  J. Chem. Phys. 82 (1985) 2795.

\bibitem{schaaf-talbot-rabeony-reiss}  P. Schaaf, J. Talbot,
           H. M. Rabeony, and H. Reiss, J. Phys. Chem. 92 (1988) 4826.

\bibitem{widom} B. Widom, J. Chem. Phys. 44 (1966) 3888; 58 (1973) 4043.

\bibitem{schaaf-talbot} P. Schaaf and J. Talbot, 
                     Phys. Rev. Lett. 62 (1989) 175.

\bibitem{hoffman} D. K. Hoffman, J. Chem. Phys. 65 (1976) 95.

\bibitem{evans-series} J. W. Evans, Physica A 123 (1984) 297; 
                    J. Chem. Phys. 87 (1989) 3038;
                    Phys. Rev. Lett. 62 (1989) 2624.

\bibitem{baram-kutasov} A. Baram and D. Kutasov, 
               J. Phys. A: Math. Gen. 22 (1989) L251.

\bibitem{dickman-wang-jensen} R. Dickman, J.-S. Wang, and I. Jensen,
                  J. Chem. Phys. 94 (1991) 8252.

\bibitem{bonnier-series} B. Bonnier, M. Hontebeyrie, and 
               C. Meyers, Physica A, 198 (1993) 1. 

\bibitem{baram-fixman} A. Baram and M. Fixman, J. Chem. Phys. 103 (1995) 1929.

\bibitem{gan-wang-JCP} C. K. Gan and J.-S. Wang, 
          J. Chem. Phys. 108 (1998) 3010. 

\bibitem{vette} K. J. Vette, T. W. Orent, D. K. Hoffman, 
                and R. S. Hansen, J. Chem. Phys. 60 (1974) 4854.

\bibitem{gan-wang-PhysA} C. K. Gan and J.-S. Wang, 
                      J. Phys. A: Math. Gen. 29 (1996) L177. 

\bibitem{cormen} T. H. Cormen, C. E. Leiserson, and R. L. Rivest,
                   ``Introduction to Algorithms,'' the MIT Press, 
                     Cambridge, 1990.

\bibitem{gan-wang-PRE} C. K. Gan and J.-S. Wang, 
                       Phys. Rev. E. 55 (1997) 107. 

\bibitem{wang-gan-PRE}  J.-S. Wang and C. K. Gan, 
                        Phys. Rev. E. 57 (1998) 6548. 

\bibitem{hansen} J. P. Hansen and I. R. McDonald,
                  ``Theory of Simple Liquids,'' Academic Press,
                     London, 1976.

\bibitem{mayer-theory} G. E. Uhlenbeck and G. E. Ford, in
                     ``Studies in Statistical Mechanics,''
                      Part B, Vol. 1,
                      edited by J. de Boer and G. E. Uhlenbeck
                      North-Holland, Amsterdam, 1962;
                      H. L. Friedman, ``A Course in Statistical
                      Mechanics,'' Chapter 6, Prentice-Hall,
                      Englewood Cliffs, 1985.

\bibitem{given} J. A. Given, Phys. Rev. A 45 (1992) 816.

\bibitem{tarjus-schaaf-talbot-JSP} G. Tarjus, P. Schaaf, and
                   J. Talbot, J. Stat. Phys. 63 (1991) 167.

\bibitem{hoover} W. G. Hoover and A. G. De Rocco, 
                 J. Chem. Phys. 36 (1962) 3141. 

\bibitem{geng} B. D. McKay, J. Algorithms, 26 (1998) 306.

\bibitem{reiss-frisch-lebowitz} H. Reiss, H. L. Frisch, and
           J. L. Lebowitz, J. Chem. Phys. 31 (1959) 369. 

\bibitem{schaaf-talbot-JCP} P. Schaaf and J. Talbot,
                      J. Chem. Phys. 91 (1989) 4401.

\bibitem{talbot-schaaf-tarjus-MP} J. Talbot, P. Schaaf, and 
                   G. Tarjus, Mol. Phys. 72 (1991) 1397.

\bibitem{ricci-etal} S. M. Ricci, J. Talbot, G. Tarjus, and P. Viot,
                    J. Chem. Phys. 97 (1992) 5219. 

\bibitem{caser-hilhorst}  S. Caser and H. J. Hilhorst,
              J. Phys. A: Math. Gen. 27 (1994) 7969;
              J. Phys. A: Math. Gen. 28 (1995) 3887.
               
\bibitem{fan-percus} Y. Fan and J. K. Percus, 
                     Phys. Rev. Lett.  67 (1991) 1677;
                     Phys. Rev. A, 44 (1991) 5099.

\bibitem{levin} D. Levin, J. Comput. Math. 3 (1973) 371.

\bibitem{meakin-cardy-etal} P. Meakin, J. L. Cardy, E. Loh, Jr.,
                    and D. J. Scalapino, J. Chem. Phys.
                    86 (1987) 2380.

\bibitem{wang-pandey} J.-S. Wang and R. B. Pandey, 
                     Phys. Rev. Lett. 77 (1996) 1773. 

\bibitem{nakamura} M. Nakamura, J. Phys. A: Math. Gen. 
                   19 (1986) 2345.

\bibitem{barker-grimson} G. C. Barker and M. J. Grimson, 
                         Mol. Phys. 63 (1988) 145;
                         J. Phys. A: Math. Gen. 20 (1987) 2225.

\bibitem{privman-wang-nielaba} V. Privman, J.-S. Wang, and P. Nielaba,
                    Phys. Rev. B, 43 (1991) 3366. 

\bibitem{akeda-hori} Y. Akeda and M. Hori, Nature, 254 (1975) 318;
                     Biometria, 63 (1976) 361.

\bibitem{tanemura} M. Tanemura, Ann. Inst. Stat. Math. 31 (1979) 351.

\bibitem{hinrichsen-etal}  E. L. Hinrichsen, J. Feder, and T. J\o ssang, 
        J. Stat. Phys.  44 (1986) 793.

\bibitem{meakin-jullien} P. Meakin and R. Jullien, 
                   Phys. Rev. A, 46 (1992) 2029;
                   Physica A 187 (1992) 475.

\bibitem{wang-ijmp} J.-S. Wang, Int. J. Mod. Phys.  C, 5 (1994) 707. 

\bibitem{palasti} I. Pal\'asti, Magy. Tud. Akad. Mat. Kut. Int\'ez.
                  K\"ozl. 5 (1960) 353.

\bibitem{brosilow-ziff-vigil} B. J. Brosilow, R. M. Ziff, and R. D. Vigil, 
                Phys. Rev. A 43 (1991) 631.
                  
\bibitem{talbot-tarjus-schaaf-PRA} J. Talbot, G. Tarjus, and P. Schaaf,
                           Phys. Rev. A 40 (1989) 4808. 

\bibitem{sherwood} J. D. Sherwood, J. Phys. A. Math. Gen. 23 (1990) 2827.

\bibitem{ziff-vigil} R. M. Ziff and R. D. Vigil,
                     J. Phys. A: Math. Gen. 23 (1990) 5103. 

\bibitem{vigil-ziff} R. D. Vigil and R. M. Ziff, 
                    J. Chem. Phys. 91 (1989) 2599; 93 (1990) 8270.

\bibitem{viot-etal} P. Viot, G. Tarjus, S. M. Ricci, and J. Talbot,
                    J. Chem. Phys. 97 (1992) 5212.

\bibitem{adamczyk} Z. Adamczyk and P. Wero\'nski,
                   J. Chem. Phys. 105 (1996) 5562.

\bibitem{talbot-schaaf-mixture} J. Talbot and P. Schaaf,
                  Phys. Rev. A 40 (1989) 422.

\bibitem{tarjus-talbot-mixture} G. Tarjus and J. Talbot,
                  J. Phys. A: Math. Gen. 24 (1991) L913.

\bibitem{oliveira-tome-dickman} M. J. de Oliveira, T. Tom\'e, and
                R. Dickman, Phys. Rev. A, 46 (1992) 6294.

\bibitem{manna} S. S. Manna and N. M. \u Svraki\'c, 
                    J. Phys. A 24 (1991) L671.

\bibitem{svrakic-henkel} N. M. \u Svraki\'c and M. Henkel,
                J. Phys. I, 1 (1991) 791. 

\bibitem{bonnier-mixture} B. Bonnier, Europhys. Lett. 18 (1992) 297.

\bibitem{sinkovits-pandey} R. S. Sinkovits and R. B. Pandey,
                        J. Stat. Phys. 74 (1994) 457.

\bibitem{budinski} L. Budinski-Petkovi\'c and U. Kozmidis-Luburi\'c,
                   Phys. Rev. E. 56 (1997) 6904; 
                   Physica A 236 (1997) 211;  
                   Physica A 262 (1999) 388.

\bibitem{nielaba-privman-wang} P. Nielaba, V. Privman, and J.-S. Wang, 
           J. Phys. A: Math. Gen. 23 (1990) L1187; and 
in {\it Computer Simulation Studies in Condensed-Matter Physics VI},
D. P. Landau, K. K. Mon, H.-B. Sch\"uttler, eds., Springer Proceedings
in Physics, Vol.~76, p.~143, Springer-Verlag, Heidelberg, 1993.

\bibitem{hilfer-wang} R. Hilfer and J.-S. Wang, 
                   Phys. A: Math. Gen. 24 (1991) L389. 

\bibitem{privman-nielaba-europhys} V. Privman and P. Nielaba,
                            Europhys. Lett. 18 (1992) 673.

\bibitem{privman-barma} V. Privman and M. Barma,
                      J. Chem. Phys. 97 (1992) 6714.

\bibitem{wang-defect} J.-S. Wang, P. Nielaba, and V. Privman, 
               Mod. Phys. Lett. B 7 (1993) 189;
               Physica A, 199 (1993) 527. 

\bibitem{pereira} G. G. Pereira and J.-S. Wang, 
                  Physica A 242 (1997) 347.

\bibitem{bonnier} B. Bonnier, Phys. Rev. E. 56 (1997) 7304.

\bibitem{eisenberg-baram} E. Eisenberg and A. Baram,
                    Europhys. Lett. 44 (1998) 168.

\bibitem{bortz-kalos-lebowitz} A. B. Bortz, M. H. Kalos, and J. L. Lebowitz,
                      J. Comput. Phys. 17 (1975) 10. 

\end{thebibliography}
\end{document}